\documentclass[prb,aps,twocolumn,showpacs,superscriptaddress]{revtex4-1}
\usepackage{graphicx,bm,amsmath,amssymb,psfrag}

\begin{document}
\title{Energy spectra for quantum wires and 2DEGs in magnetic fields with Rashba and Dresselhaus spin-orbit interactions}
\author{Sigurdur I.\ Erlingsson}
\affiliation{School of Science and Engineering, Reykjavik University, Menntavegi 1, IS-101 Reykjavik, Iceland}
\email{sie@ru.is}
\altaffiliation[Permanent address: ]
{School of Science and Engineering, Reykjavik University, Menntavegi 1, IS-101 Reykjavik, Iceland}
\affiliation{Science Institute, University of Iceland, Dunhagi 3, IS-107
  Reykjavik, Iceland}
\author{J.\ Carlos Egues}
\affiliation{Departamento de F\'{\i}sica e Inform\'atica, Instituto de  F\'{\i}sica de  S\~{a}o  Carlos, Universidade de S\~{a}o Paulo, \\
13560-970 S\~{a}o Carlos,  S\~{a}o Paulo, Brazil}
  \affiliation{Department of Physics, University of Basel, Klingelbergstrasse 82, CH-4056 Basel,
  Switzerland}
\author{Daniel Loss}
\affiliation{Department of Physics, University of Basel, Klingelbergstrasse 82, CH-4056 Basel,
  Switzerland}
  \date{\today}
\begin{abstract}
We introduce an analytical approximation scheme to diagonalize parabolically confined two dimensional electron systems
with both the Rashba and Dresselhaus spin-orbit interactions. The starting point of our perturbative expansion is a zeroth-order Hamiltonian
for an electron confined in a quantum wire with an effective spin-orbit induced magnetic field along the wire, obtained by properly rotating the usual spin-orbit Hamiltonian.
We find that the spin-orbit-related transverse coupling terms can be recast into two parts $W$ and $V$, which
couple crossing and non-crossing adjacent transverse modes, respectively. Interestingly, the zeroth-order Hamiltonian
together with $W$ can be solved exactly, as it maps onto the Jaynes-Cummings model of quantum optics. We treat the $V$
coupling by performing a Schrieffer-Wolff transformation. This allows us to obtain an effective Hamiltonian to third
order in the coupling strength $k_\mathrm{R}\ell$ of $V$, which can be straightforwardly diagonalized via an additional
unitary transformation. We also apply our approach to other types of effective parabolic confinement, e.g., 2D electrons
in a perpendicular magnetic field.
To demonstrate the usefulness of our approximate eigensolutions, we obtain
analytical expressions for the $n^{th}$ Landau-level $g_{n}$-factors  in the presence of both
Rashba and Dresselhaus couplings. For small values of the bulk \textit{g}-factors, we find that spin-orbit effects cancel out entirely  for particular values of the spin-orbit couplings. By solving simple transcendental equations we also
obtain the band minima of a Rashba-coupled
quantum wire as a function of an external magnetic field. These can
be used to describe Shubnikov-de Haas oscillations. This procedure makes it easier
to extract the strength of the spin-orbit interaction in these systems via proper fitting
of the data.
\end{abstract}
\maketitle
\section{Introduction}
For a wide range of systems studied in quantum transport the starting point of
the sample fabrication is a two dimensional electron gas (2DEG). Various
nanostructures are defined in the 2DEG by using either metallic gates to
expel electrons or by etching into the electron gas\cite{davies07:book}.
Quantum wires can be formed using both methods and in some cases the wire
confinement can be assumed parabolic.

Since the seminal experiments showing conductance quantization\cite{vanwees91:12431,wharam88:L209} in
narrow gate-tunable constrictions acting as quantum wires (or quantum point contacts) defined in 2DEGs, these
structures have been extensively used to investigate a rich variety of physical phenomena, e.g.,
the 0.7 anomaly and analogs \cite{thomas96:135,graham03:136404}. Spin orbit interaction has also been
investigated experimentally\cite{guzenko07:577,schaepers09:06001} in parallel quantum wires, where universal conductance
fluctuations are suppressed. More recently, ballistic spin resonance
due to an intrinsically oscillating spin-orbit field has been experimentally
realized in a quantum wire\cite{frolov09:868}. The observation of the `one-dimensional spin-orbit gap' in quantum wires has also been
reported recently\cite{quay10:336}. Both of these experiments highlight the use of
spin-orbit effects in quantum wires as a means to control the spin of carriers, an important ingredient
for potential spintronic applications.

Bulk semiconductors lacking an inversion center in the crystal lattice have a built-in
spin-orbit interaction, the so called Dresselhaus term\cite{dresselhaus55:580}. Heterostructures in which
the 2DEG is formed by an asymmetric confining potential also exhibit the Rashba spin-orbit
interaction\cite{bychkov84:6039}. As shown experimentally, the Rashba coupling strength can be tuned via proper
gating of the structure\cite{nitta97:1335,engels97:1958R}, which makes the Rashba interaction very appealing for potential
technological applications involving spin control. More recently, yet a new type of spin-orbit
interaction has been found in symmetric two-dimensional quantum structures with two subbands:
the inter-subband-induced spin-orbit coupling\cite{bernardes07:076603,calsaverini08:155313}.
Interestingly, this new spin-orbit interaction gives rise to non-zero intrinsic spin Hall effect\cite{lee09:155314}.

Here we consider a parabolically confined asymmetric (narrow) quantum well with only the lowest occupied subband  and in the 
presence of both the Dresselhaus and the Rashba terms
\begin{eqnarray}
H&=&\frac{1}{2m^*}(p_x^2+p_y^2)+ \frac{\alpha}{\hbar}(p_y \sigma_x-p_x\sigma_y) + \nonumber  \\
& & \frac{\beta}{\hbar}(p_x \sigma_x-p_y \sigma_y) +\frac{1}{2}m^*
\omega_0^2y^2,
\label{eq:Hso}
\end{eqnarray}
where $m^*$ is the electron effective mass, $\omega_0$ characterizes the strength of the parabolic confinement, $p_{x(y)}$ is the
momentum operator in the ${x(y)}$ direction, and $\alpha $ and $\beta$ are the Rashba and
the Dresselhaus coupling strengths, respectively. Similar systems have been studied before in the presence of Landau level quantization,
using either variational (Hartree-Fock) methods \cite{schliemann03:085302}, second order
perturbation \cite{zarea05:085342,wang05:085344} or by obtaining the
energy spectrum in terms of perturbation series \cite{zhang06:L477}.  However, we are unaware of any analytical
solution for both Rashba and Dresselhaus that goes beyond second order perturbation for multiple levels, i.e.,
\ not just treating the two lowest orbital levels as in Ref.\ \cite{perroni07:186227}.)
We note that in the absence of magnetic fields the spectrum of
2D electrons with both the (linear-in-momentum) Rashba and Dresselhaus
interactions is known; for non-zero magnetic fields only particular cases are
solvable but the general solution is not known. The above model Hamiltonian has also been investigated in connection with the
Zitterbewegung of injected spin-polarized wave packets in quantum wells and wires.\cite{schliemann06:085323,schliemann05:206801}

We have developed a perturbative scheme to diagonalize our Hamiltonian Eq. (\ref{eq:Hso}) based on
separating it into terms that can be treated exactly (``effective Jaynes-Cummings model'') plus a perturbative term .
In Section \ref{sec:Rashba} we derive the perturbative scheme in the case of a parabolic quantum wire
with Rashba spin-orbit coupling and in Section \ref{sec:RashbaDresselhaus} we extend it
to include both the Rashba and Dresselhaus interactions.
These results are then applied in Section \ref{sec:magnetotransport} to magnetotransport in
(i) quantum wires with Rashba coupling and (ii) 2DEGs in a perpendicular magnetic field in
the presence of both Rashba and Dresselhaus couplings. Both of these systems can be mapped
onto the Hamiltonian in Eq. (1) and the general analytical
results derived straightforwardly used. For (i) the approximate eigenvalues
allow us to obtain beating patterns that directly relate to the Shubnikov-de Haas oscillations,
by simply finding the roots of transcendental equations.  This can be used to more easily extract the Rasbha
coupling strength from experimental data \cite{guzenko07:577}. For (ii) we derive an expression for
the effective $g$-factor of Landau levels.  Interestingly, we find that when the $g$-factor is small enough,
the spin-orbit effects cancel out exactly for certain values of the Rashba and Dresselhaus couplings.

\section{Pure Rashba spin-orbit coupling}
\label{sec:Rashba}
When only the Rashba term is present the Hamiltonian of the system is
\begin{eqnarray}
H_\mathrm{R}=\frac{1}{2m^*}(p_x^2+p_y^2)+\frac{1}{2}m^*
\omega_0^2y^2+\frac{\alpha}{\hbar}(p_y \sigma_x-p_x\sigma_y).
\end{eqnarray}
It is convenient to introduce the standard ladder
operator which yields the new Hamiltonian
\begin{eqnarray}
H_\mathrm{R}=\frac{p_x^2}{2m^*}+\hbar \omega_0 a^\dagger a-
\frac{\hbar k_\mathrm{R}p_x}{m^*} \sigma_y+\frac{i\hbar^2 k_\mathrm{R}}{\sqrt{2}m^*\ell}(a^\dagger-a)\sigma_x,
\label{eq:HRorig}
\end{eqnarray}
where we have introduced the Rashba wave vector $k_\mathrm{R}=\frac{m^*\alpha}{\hbar^2}$, the oscillator length $\ell=\sqrt{\hbar/m^*\omega_0}$, 
and the energy is measured relative to $\hbar \omega_0/2$.
The  system is translationally invariant along the wire (i.e., $[p_x,H_R]=0)$ so we seek
eigensolutions of $H_\mathrm{R}$ which are plane waves in this direction and hence
eigenvectors of $p_x$ with eigenvalues $\hbar k$,
\begin{eqnarray}
H_\mathrm{R}&=&\frac{1}{2}k^2+a^\dagger a-k_\mathrm{R} k \sigma_y
+\frac{ik_\mathrm{R}}{\sqrt{2}}(a^\dagger-a)\sigma_x.
\label{eq:HRscaled}
\end{eqnarray}
The energy in Eq.\ (\ref{eq:HRscaled}) is measured in units of $\hbar \omega_0$,
and momenta (both $\hbar k$ and $\hbar k_\mathrm{R}$) in $l^{-1}$.
At a first glance, the above Hamiltonian looks like a shifted harmonic oscillator; however,
since $\sigma_x$ and $\sigma_y$ do not commute, an exact solution is not
known.

\begin{figure}[t]
\begin{center}
\psfrag{EAxis}{\Huge $\varepsilon^0_{ns}(k)-\frac{1}{2}k^2$}
\psfrag{kAxis}{\Huge $k/k^*$}
\psfrag{spinUp}{\Huge $s=+1$}
\psfrag{spinDown}{\Huge $s=-1$}
\includegraphics[angle=-90,width=8.7cm]{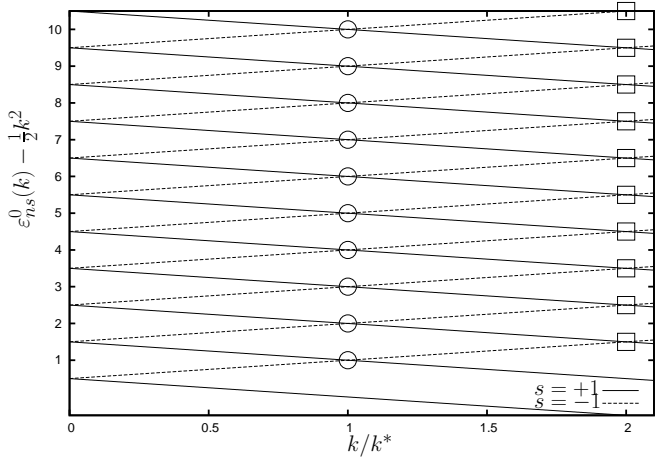}
\caption{Energy spectrum of the zeroth-order Hamiltonian $\tilde{H}_0-\frac{1}{2}k^2$ [Eq. (\ref{eq:H0tilde})]
for $k_\mathrm{R}=0.25$ (first ten wire modes shown). The circles and
squares indicate the crossings between adjacent and
next-to-adjacent, respectively, opposite-spin oscillator states. Here $k^* =\frac{1}{2k_\mathrm{R}}$ defines
the crossing point of the energy dispersions of the states $|k,n,-1\rangle$ and $|k,n+1,+1\rangle$.}
\label{fig:zeroDisp}
\end{center}
\end{figure}

Our goal is to construct a perturbation expansion in the small parameter
$k_\mathrm{R}$ but note that the product $k_\mathrm{R} k$ need not be
small.
It is convenient to rotate the spin operators to obtain a new
Hamiltonian where the effective magnetic field due to the momentum along the
wire couples to $\sigma_z$
\begin{eqnarray}
\tilde{H}_\mathrm{R}&=&e^{-i\frac{\pi}{4}\sigma_x}H_\mathrm{R}e^{i\frac{\pi}{4}
\sigma_x}
\nonumber \\
&=&\frac{1}{2}k^2+a^\dagger a-k_\mathrm{R} k \sigma_z
+\frac{ik_\mathrm{R}}{\sqrt{8}} (\sigma_+(a^\dagger-a)  -\mbox{h.c. } ), \nonumber \\
\label{eq:HRtilde}
\end{eqnarray}
where $\sigma_+=\sigma_x+i\sigma_y$.
Important features of the spectrum and the strategy of the following
perturbative expansion are best explained in terms of the zeroth-order Hamiltonian
\begin{eqnarray}
\tilde{H}_0=\frac{1}{2}k^2+a^\dagger a-k_\mathrm{R} k\sigma_z.
\label{eq:H0tilde}
\end{eqnarray}
The kets $|k,n,s\rangle$ are eigenstates of $\tilde{H}_0$. These are also simultaneous
eigenstates of $a^\dagger a$ and $\sigma_z$ with eigenvalues $n=0,1,\dots$ and
$s=\pm 1$, respectively.
The eigenvalues $\varepsilon^0_{ns}(k)-\frac{1}{2}k^2=n-sk_\mathrm{R} k$ are
plotted in Fig.\ \ref{fig:zeroDisp} for $k_\mathrm{R}=0.25$.  Focusing on
positive wavevectors $k>0$ one sees that the state $|k,n,-1\rangle$ crosses the
state $|k,n+1,+1\rangle$ at  $k^* =\frac{1}{2k_\mathrm{R}}$.
This means that non-degenerate perturbation theory cannot be applied to the
terms containing $a^\dagger \sigma_+$ (or their Hermitian conjugate) as they couple these states.  On the
other hand, states $|k,n,+1\rangle$ never cross  $|k,n+1,-1\rangle$ and thus the
coupling terms $a^\dagger \sigma_-$, and their hermitian conjugate, can be
treated perturbatively.
The negative wavevector part of the spectrum is analyzed in the same manner using
the Kramers relation $\varepsilon^0_{ns}(-k)=\varepsilon^0_{n,-s}(k)$, i.e., for
$k<0$ the spin indices in the above discussion are reversed.

The above considerations motivate us to separate the transverse mode coupling into
two parts $W$ and $V$
\begin{eqnarray}
W&=&\frac{i k_\mathrm{R}}{\sqrt{8}}(a^\dagger \sigma_+ - a\sigma_-) \label{eq:W} ,  \\
V&=&\frac{i k_\mathrm{R} }{\sqrt{8}}(a^\dagger \sigma_- - a\sigma_+),
\label{eq:V}
\end{eqnarray}
where $W$ contains the term that couple states that cross and $V$ is the part of
the coupling that can be treated pertubatively (i.e., $V$ does not couple
states that cross). It is worth pointing out that $V$ and $W$ are related via time reversal symmetry, i.e.,
$V=\mathcal{T}^{-1} W\mathcal{T}$, where $\mathcal{T}$ is the time reversal
operator \cite{sakurai85:xx}, i.e.\ their roles are reversed for the negative
$k$ states. The following approximation scheme does not rest upon time reversal symmetry.
Time reversal is only used to obtain the negative wavevector ($k<0$) states from
the positive ones.

Thus we split the Hamiltonian in Eq. (\ref{eq:HRtilde}) into two parts:
\begin{eqnarray}
 \tilde{H}_\mathrm{R}&=&\tilde{H}_{\mathrm{R},JC}+V,
\label{eq:HRtilde-Sep}
\end{eqnarray}
where $V$ is defined in Eq.\ (\ref{eq:V}) and $\tilde{H}_{\mathrm{R},JC}$ is a Jaynes-Cummings-like Hamiltonian \cite{jaynes63:89,zuelicke07:355}
\begin{eqnarray}
\tilde{H}_{\mathrm{R},JC} &=& \tilde{H}_0+W \nonumber  \\
&=&\frac{1}{2}k^2+a^\dagger a-k_\mathrm{R} k \sigma_z +\frac{i
k_\mathrm{R}}{\sqrt{8}}(a^\dagger \sigma_+ -
a\sigma_-), \nonumber \\
\label{eq:HRJC}
\end{eqnarray}
which is exactly diagonalizable. In fact, we show in Appendix \ref{app:rotation}
that a generalized coupling of the form
$[\gamma(\hat{n})a^\dagger \sigma_+ + \mbox{h.c.}]$, with $\gamma$ being some generic complex function of $\hat{n}=a^\dagger a $, can be diagonalized.
This will guide us in determining which terms to keep and which to discard in the following perturbative expansion in $V$.

\subsection{Effective Rashba wire Hamiltonian}

We start with defining a new Hamiltonian using the transformation
\begin{eqnarray}
H_\mathrm{R}'&=& e^{S}\tilde{H}_\mathrm{R}e^{-S},
\label{eq:HRprimeDefinition}
\end{eqnarray}
where $S$ is chosen such that
\begin{eqnarray}
[S,\tilde{H}_0+W]+V=0+\mathcal{O}(k_\mathrm{R}^4).
\label{eq:Sdefinition}
\end{eqnarray}
This procedure is the same as the usual Schrieffer-Wolff transformation\footnote{Let $S= S_1+S_2+S_3$,
where the first order term $S_1$ is determined by $[S_1,H_0]+V=0$.  The second
order term $S_2$ is determined by the condition $[S_1,W]+[S_2,H_0]=0$.  This in turn gives an equation
for $S_3$: $[S_2,W]+[S_3,H_0]=0$.}, the only difference is that the
zeroth-order Hamiltonian in Eq.\ (\ref{eq:HRJC}) contains a term proportional to the perturbative
parameter $k_\mathrm{R}$.\cite{bruus04:book}. We find that
\begin{eqnarray}
S&=&
\left (
\frac{-ik_\mathrm{R} a^\dagger \sigma_-}{\sqrt{8}(1 +2k_\mathrm{R} k)} +
\frac{k_\mathrm{R}^2\sigma_z
  {a^\dagger}^2}{4(1 +2 k_\mathrm{R} k)} \right . \nonumber \\
& & \left .
k_\mathrm{R}^3 \mathcal{A}_k  a \sigma_+ a^\dagger a +\frac{k_\mathrm{R}^3 \mathcal{B}_k}{3-2k_\mathrm{R} k}a^\dagger \sigma_+ \right )- \mbox{h.c.}.
\label{eq:S}
\end{eqnarray}
In the last line we have introduced the factors $\mathcal{A}_k$ and $\mathcal{B}_k$ which  depend only on $k$.
Here we do not write out the explicit forms of $\mathcal{A}_k$ and
$\mathcal{B}_k$ since they will only occur in terms in the effective Hamiltoninan $\propto  k_\mathrm{R}^4$
(the extra power of $k_\mathrm{R}$ coming from $[S,V]$) and can thus
be discarded when certain restrictions on $k$ are taken into account.
The denominator in the $\mathcal{B}_k$-term vanishes when $k=3k^*$,
since at that $k$-value it couples states that are degenerate.
In order to treat this term perturbatively we make the restriction $| 3k^*-k|
\gtrsim  k_\mathrm{R}$.

We can then proceed with the usual calculations, i.e.\ knowing the form of $S$
the transformed Hamiltonian is
\begin{eqnarray}
H_\mathrm{R}'&=&\tilde{H}_0+W+\frac{1}{2}[S,V]+
   \frac{1}{3}[S,[S,V]] \nonumber  \\
& &+\mathcal{O}(k_\mathrm{R}^4). \label{eq:HRprimeExpansion}
\end{eqnarray}
Inserting Eq.\ (\ref{eq:S}) into Eq.\ (\ref{eq:HRprimeExpansion}) we obtain the
following exact result (up to the order indicated) for the Hamiltonian
\begin{eqnarray}
H_\mathrm{R}'&=&\tilde{H}_0+W+k_\mathrm{R}^2 \frac{1-\tau_z(2a^\dagger a +1)}{4
  (1+2 k_\mathrm{R} k)} \nonumber \\
& &+k_\mathrm{R}^3 \frac{-i}{8\sqrt{2}(1+2k_\mathrm{R}
  k)} ( a^\dagger a a^\dagger \sigma_+- \mbox{h.c.}) \nonumber \\
& &+k_\mathrm{R}^3  \mathcal{C}_k\left [ \left (\frac{1}{3}  a \sigma_+ a^\dagger a+
  \frac{1}{8}{a^\dagger}^3\sigma_- \right ) +\mbox{h.c.} \right ],
\label{eq:HRprime}
\end{eqnarray}
where $\mathcal{C}_k$ is a $k$-dependent factor.  Note that the term proportional to $\mathcal{C}_k$ only contains operators that
couple states separated in energy by {\em at least} $\hbar \omega$ and result in corrections in eigenenergies $\propto k_\mathrm{R}^6$ and can
thus be dropped.  This results in the effective Hamiltonian for the Rashba wire
\begin{eqnarray}
H_\mathrm{R,eff}&=&
\frac{k^2}{2}+\hat{n}-k_\mathrm{R} k \,\sigma_z -\frac{k_\mathrm{R}^2}{4}
\frac{1+(2\hat{n} +1)\sigma_z}{(1+2 k_\mathrm{R} k)}  \nonumber \\
& &+ \frac{1}{2}\bigl [\gamma a^\dagger \sigma_+ +a \sigma_-\gamma^* \bigr ],
\label{eq:HReff}
\end{eqnarray}
where the function $\gamma$ is defined as
\begin{eqnarray}
\gamma \equiv \gamma(\hat{n})&=&\frac{-i k_\mathrm{R}}{\sqrt{2}}
 \left (
1-\frac{k_\mathrm{R}^2 \hat{n}}{4 (1 +2k_\mathrm{R} k)}  \right  ).
\label{eq:gammaR}
\end{eqnarray}
The order of the operators in the last term of Eq.\ (\ref{eq:HReff}) is important
since $[a,\gamma]\neq 0$.

\subsection{Diagonalizing $H_\mathrm{R,eff}$}

An interesting feature of $H_\mathrm{R,eff}$, Eq.\ (\ref{eq:HReff}), is that it can be diagonalized via a generalized rotation 
matrix $R(k)$ (see Appendix \ref{app:rotation}), which results in
\begin{eqnarray}
H_{\mathrm{R,diag}}&=&R^\dagger H_{\mathrm{R,eff}}R \nonumber \\
&=& \left (
\begin{array}{cc}
H_+(\hat{n}) & 0 \\
0 & H_-(\hat{n})
\end{array}
\right ),
\label{eq:HReffRot}
\end{eqnarray}
where $H_\pm(\hat{n})$ are given in Appendix \ref{app:rotation}.
The eigenstates of $H_{\mathrm{R,diag}}(k)$ are denoted by $|k,n,s \rangle$ and corresponding the eigenenergies
$\varepsilon_{ns}(k)$ are given by
\begin{eqnarray}
\varepsilon_{n,\uparrow}(k)&=&
\frac{k^2}{2}+n-\frac{k_\mathrm{R}^2/2}{1+2k_\mathrm{R} k}
+\Delta_{n}(k) ,
\label{eq:EnU}\\
\varepsilon_{n,\downarrow}(k)&=&
\frac{k^2}{2}+n+1-\frac{k_\mathrm{R}^2/2}{1+2k_\mathrm{R} k}
-\Delta_{n+1}(k).
\label{eq:EnD}
\end{eqnarray}
The generalized Jaynes-Cummings coupling mixes adjacent transverse states with opposite spin, that
cross at $k^*$ (indicated by the empty circles in Fig.\ \ref{fig:zeroDisp}).  This mixing is described by $\Delta_n(k)$
\begin{eqnarray}
\Delta_n(k)&=&\frac{1}{2}\left [ \left ( 1 -2k_\mathrm{R}  k-\frac{k_\mathrm{R}^2
n}{1+2k_\mathrm{R}  k} \right )^2 \right . \nonumber \\
 & & \left .+ 2 k_\mathrm{R}^2 n\left ( 1-\frac{k_\mathrm{R}^2 n}{4(1+2
   k_\mathrm{R} k)}\right )^2 \right ]^{1/2}.
\label{eq:Delta}
\end{eqnarray}
Note that the $(n,s)=(0,\uparrow)$ state reduces to
\begin{eqnarray}
\varepsilon_{0,\uparrow}(k)&=&
\frac{k^2}{2}-k_\mathrm{R} k
-\frac{k_\mathrm{R}^2}{2(1+2k_\mathrm{R}  k)},
\label{eq:E0U}
\end{eqnarray}
which reflects the fact that $|0,\uparrow \rangle$ is an eigenstate of the
Hamiltonian in Eq.\ (\ref{eq:HReff}).

The above perturbative scheme does not depend on the fact that both $W$ and $V$ have equal coupling strengths.  For completeness we give the
eigenvalues for different coupling values in Appendix \ref{app:unequalVW}.
\begin{figure}[b]
\begin{center}
\psfrag{EAxis}{\Huge $\varepsilon_{ns}(k)-\frac{1}{2}k^2$}
\psfrag{kAxis}{\Huge $k/k^*$}
\psfrag{spinUp}{\Huge $s=+1$}
\psfrag{spinDown}{\Huge $s=-1$}
\psfrag{nEq9}{\Huge $n=9$}
\psfrag{nEq0}{\Huge $n=0$}
\includegraphics[angle=-90,width=8.7cm]{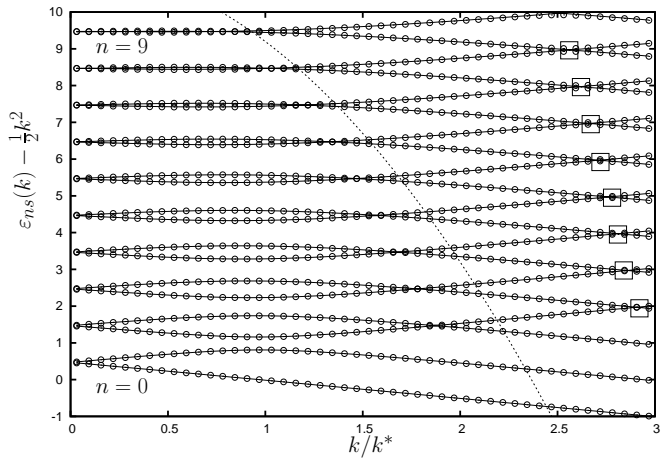}
\caption{Eigenenergies of the first ten Rashba wire modes for $k_\mathrm{R}=0.25$ (which yields $k^*=2$). 
The solid lines are obtained within our analytical approximation scheme [Eqs. (\ref{eq:EnU}) and (\ref{eq:EnD})], while the empty circles 
correspond to the (exact) eigenenergies from a numerical diagonalization of $H_\mathrm{R}$ in Eq. (\ref{eq:HRscaled}). 
The deviation between the numerical and the approximate analytical spectra cannot be seen on scale of the plot. 
The dashed line denotes the Fermi energy $E_F$ and the empty squares show the crossings of the states separated by $3\hbar \omega_0$.}
\label{fig:dispRashba025}
\end{center}
\end{figure}

\subsection{Numerical vs analytical results: Rashba case}

To better understand the range of validity of the above approximation scheme, it is
instructive to look at the original Hamiltonian in Eq.\ (\ref{eq:HRscaled}).  When $k=0$ the Hamiltonian is easily
diagonalized since it is just a shifted harmonic oscillator.  The eigenstates of
$H_\mathrm{R}(0)$ are $e^{\frac{i\pi}{4}\sigma_y}|n,s \rangle$, here $|0,n,s \rangle=|n,s \rangle$,  with eigenvalues
\begin{eqnarray}
\varepsilon_{n,s}(0)&=& n-\frac{k_\mathrm{R}^2}{2}.
\label{eq:spectrumZero}
\end{eqnarray}
Note that only the eigenvalues in Eq.\ (\ref{eq:E0U}) reduce to the exact solution in Eq.\
(\ref{eq:spectrumZero}), those in Eqs.\ (\ref{eq:EnU}) - (\ref{eq:EnD})
do not.  The magnitude of this deviation can be used to determine the accuracy of the
analytical spectum.  The origin of this deviation is that we have chosen to find an
approximation scheme valid for $k>0$, but not at $k=0$.  This allowed us to do
the necessary approximations to obtain (\ref{eq:HReff}), whose maximum
deviation from the exact results occurs at $k=0$.
Comparing Eqs.\ (\ref{eq:EnD})-(\ref{eq:E0U}) to Eq.\ (\ref{eq:spectrumZero})
gives the deviation of the approximate solution and the exact one at $k=0$.
Requiring that the deviation of the approximate solution and the exact solution
be much smaller than the energy separation of the transverse states
\begin{eqnarray}
\Delta \varepsilon&=&|\varepsilon_{ns}(0)-(n-k_\mathrm{R}^2/2)| \ll 1,
\label{eq:accuracyRequirement}
\end{eqnarray}
yields the inequality $\frac{k_\mathrm{R}^2 n}{2} \ll 1$,
where we have expanded the square root in Eq.\ (\ref{eq:Delta}) in powers of $k_\mathrm{R}$.
Fixing the required value of the absolute accuracy $\delta$,
e.g., $\delta=5\times 10^{-2}$ for a $5\%$ accuracy, determines the
number of transverse states that satisfy Eq.\ (\ref{eq:accuracyRequirement}).
The maximum value of $n$ is
\begin{eqnarray}
n \lesssim n_{\mathrm{max}}\equiv \frac{2(4\delta)^{1/3}}{k_\mathrm{R}^2} ,
\end{eqnarray}
which gives, e.g.\  $n_{\mathrm{max}} = 18 \,(116)$ for $k_\mathrm{R}=0.25 \,(0.1)$
and $\delta=0.05$.
In practice, this requirement is not very
restrictive since in most cases $k_\mathrm{R}$ can be quite small
and the approximate solutions are good for a very large number of transverse
modes. More specifically, the values of $\alpha$ in usual III-V semiconductors range from
2 to 20\,meV \,nm, which means that the above perturbation scheme applies to wires with
 $ \ell \lesssim  1$\,$\mu$m and $\ell \lesssim 100$\,nm, respectively.
\begin{figure}[t]
\begin{center}
\psfrag{EAxis}{\Huge $\varepsilon_{ns}(k)-\frac{1}{2}k^2$}
\psfrag{kAxis}{\Huge $k/k^*$}
\psfrag{spinUp}{\Huge $s=+1$}
\psfrag{spinDown}{\Huge $s=-1$}
\psfrag{nEq50}{\Huge $n=50$}
\psfrag{nEq59}{\Huge $n=59$}
\includegraphics[angle=-90,width=8.7cm]{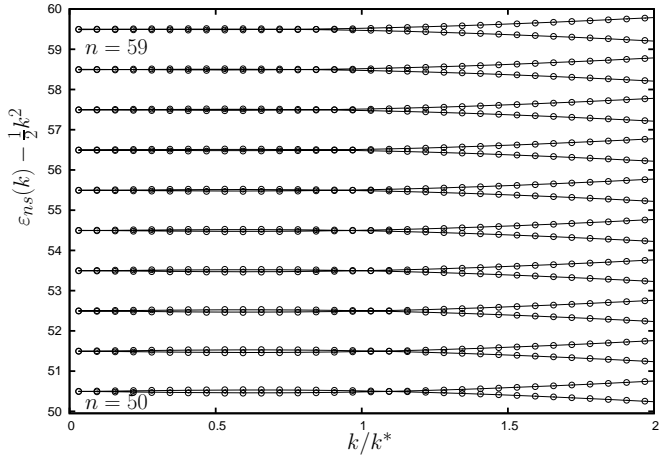}
\caption{Similar to Fig. \ref{fig:dispRashba025} but for the transverse 
states corresponding to $n=50-59$. Here the (Rashba) spin-orbit coupling strength 
$k_\mathrm{R}=0.10 \Rightarrow k^*= 5$. Note that even for such
large $n$'s the deviation of the analytical solution (lines) from the numerical
(circles) can hardly be distinguished on the scale of the plot.}
\label{fig:dispRashba010}
\end{center}
\end{figure}

Since the spin-orbit coupling is usually a small effect the contribution of the
kinetic energy $k^2/2$ is dominant in $\varepsilon_{ns}(k)$.
To emphasize the details of
the spin-orbit coupling we plot $\varepsilon_{ns}(k)-k^2/2$.
The analytical solution of the spectrum (solid lines) and numerical solutions
(circles) for different values of the coupling strength are presented in Figs.\
\ref{fig:dispRashba025} and \ref{fig:dispRashba010}.
In Fig. \ref{fig:dispRashba025} the dispersion is plotted for $k$ between $0$ and $3k^*$.  The dotted line
shows the position of the maximum value of the Fermi energy which we arbitrarily choose as $E_F=0.8(n_\mathrm{max}+1/2)$, in order
to insure that the error of approximation is less that $\delta$.
The crossing of opposite spin states separated by $\approx 3 \hbar \omega_0$ starting at
$k\approx 3k^*$ are always far above $E_F$. This shows that the effect of discarding the third order coupling term does not 
affect the accuracy of the spectrum. The same is true for all other plots that follow and in those we restrict the value of $k$ below $2k^*$.

The transverse coupling mixes adjacent states with opposite spin leading to a
family of anticrossings, as seen in Fig.\ \ref{fig:dispRashba025}. This can be
compared with  Fig.\ \ref{fig:zeroDisp} which shows the spectrum for
$k_\mathrm{R}=0.25$ and no transverse coupling.  Note how the
anticrossing at $k^*$ opens up for higher $n$ leading to effectively flat bands for the low $k$ part of the
spectrum.  This can be seen even more clearly in  Fig.\
\ref{fig:dispRashba010} for $n=50-59$ and $k_{R}=0.1$, where the dispersion is very flat.
This will result in the minima of $\varepsilon_{ns}(k)$ moving from approximately $k\approx k_\mathrm{R}$ to
$k\approx 0$ for higher values of $n$.

\section{Rashba and Dresselhaus couplings}
\label{sec:RashbaDresselhaus}
In addition to the Rashba spin-orbit coupling, there is also the Dresselhaus spin-orbit  coupling present in quantum heterostructures 
formed from semiconductors having bulk inversion asymmetry. The Dresselhaus spin-orbit  coupling is
present in III-V and II-VI material compounds, although its relative strength as compared
to the Rashba coupling can vary. Using the same units of energy and length as in Eq.\ (\ref{eq:HRscaled}) the
Rashba and Dresselhaus spin-orbit couplings (see Eq.\ (\ref{eq:Hso})) in our quantum wire can be written as
\begin{eqnarray}
H_\mathrm{RD}&=&\frac{1}{2}k^2+a^\dagger a
-(k_\mathrm{R}\sigma_y-k_\mathrm{D}\sigma_x) \,k \nonumber \\
& &+\frac{i}{\sqrt{2}}(a^\dagger -a)(k_\mathrm{R} \sigma_x-k_\mathrm{D}
\sigma_y),
\label{eq:HRDscaled}
\end{eqnarray}
where $k_\mathrm{D}$ is similar to $k_\mathrm{R}$ with $\beta$ instead of $\alpha$. The presence of the Dresselhaus
term leads to extra terms which need to be taken care of before the proceedure of the
previous section can be applied.
In the next section the $|\alpha| \gg |\beta|$ case will be treated and in the
subsequent section we will show how the $|\alpha| \ll |\beta|$ and $|\alpha|
\approx |\beta|$ cases can be written in the same form as used in the $|\alpha|
\gg |\beta|$ perturbation calculations.
\subsection{Couplings $|\alpha| \gg |\beta|$}

Let us introduce the angle $\cos\theta \equiv
k_\mathrm{R}/\sqrt{k_\mathrm{R}^2+k_\mathrm{D}^2}$ and the rotation operation
\begin{eqnarray}
k_\mathrm{R} \sigma_y-k_\mathrm{D} \sigma_x
&=&\sqrt{k_\mathrm{R}^2+k_\mathrm{D}^2} (\sigma_y\cos\theta - \sigma_x\sin
  \theta ) \nonumber \\
&=&k_\mathrm{RD}
e^{\frac{i}{2}\theta \sigma_z}\sigma_y e^{-\frac{i}{2}\theta \sigma_z},
\end{eqnarray}
where $k_\mathrm{RD}=\sqrt{k_\mathrm{R}^2+k_\mathrm{D}^2}$ is the effective spin-orbit  coupling.
Applying this spin rotation to the Hamiltonian in Eq.\ (\ref{eq:HRDscaled}), and
then the rotation used in Eq.\ (\ref{eq:HRscaled}) that takes
$(\sigma_x,\sigma_y)\rightarrow (\sigma_z,\sigma_y)$, results in the
Hamiltonian
\begin{eqnarray}
\bar{H}_{\mathrm{RD}}&\equiv&   e^{-i\frac{\pi}{4}\sigma_x}e^{-\frac{i}{2}\theta
\sigma_z}
H_\mathrm{RD}e^{\frac{i}{2}\theta \sigma_z}e^{i\frac{\pi}{4}\sigma_x}
\label{eq:HRDbarTransform} \nonumber \\
&=& \frac{1}{2}k^2+a^\dagger a
-k_\mathrm{RD} k \, \sigma_z  \nonumber \\
& &+\frac{i k_\mathrm{RD}}{\sqrt{2}}( \cos 2 \theta \sigma_x  -\sin 2\theta
\sigma_z)(a^\dagger -a).
\label{eq:HRDbar}
\end{eqnarray}
Note that the coupling becomes diagonal in $\sigma_z$ when $\cos2\theta=0$.
This occurs for values of the spin-orbit angle $\theta=\pm \frac{\pi}{4}$,
which corresponds to the case $\alpha=\pm \beta$.  In that case the exact
spectrum reduces to shifted parabolas.
The term proportional to $\sin 2\theta$ can be removed via the unitary
transformation
\begin{eqnarray}
U_z=\exp \{ -ik_\mathrm{s} \hat{x} \sigma_z\},
\end{eqnarray}
where $k_\mathrm{s}=k_\mathrm{RD}\sin 2\theta$ and $\hat{x}=\frac{1}{\sqrt{2}}(a+a^\dagger)$.
The result of applying the transformation to the Hamiltonian
$\bar{H}_\mathrm{RD}$ is
\begin{eqnarray}
\tilde{H}_\mathrm{RD}&=&
U_z^\dagger  \bar{H}_\mathrm{RD}  U_z \nonumber \\
&=& \frac{1}{2}k^2+a^\dagger a
-k_\mathrm{RD} k \, \sigma_z  -\frac{1}{2}(k_\mathrm{s})^2
\nonumber \\
& &\!\!\! +\frac{i k_\mathrm{c}}{\sqrt{8}}
 \left ( \sigma_+ e^{-ik_\mathrm{s}\hat{x}}(a^\dagger
-a)e^{-ik_\mathrm{s}\hat{x}}
-\mbox{h.c.\ } \right ). \nonumber  \\
\label{eq:HRDtilde}
\end{eqnarray}
Here $k_\mathrm{c}=k_\mathrm{RD} \cos 2\theta$, along with $k_\mathrm{s}$, will play the role of perturbation parameters as
$k_\mathrm{R}$ did in Eq.\ (\ref{eq:HRtilde}). The above Hamiltonian has a
similar structure to that in Eq. \ (\ref{eq:HRtilde}), being identical to it when $k_\mathrm{s}=0$, i.e., when
$\theta=0,\pm\frac{\pi}{2}$.  The latter two values correspond to a pure
Dresselhaus $\beta=\mp |\alpha|$, where the sign is determined by the
requirement of the perturbation procedure: The velocity is positive for $s=-1$
states at $k=0$, see Fig.\ \ref{fig:zeroDisp}.
\begin{figure}[b]
\begin{center}
\psfrag{EAxis}{\Huge $\mathcal{E}_{ns}(k)-\frac{1}{2}k^2$}
\psfrag{kAxis}{\Huge $k/k^*$}
\psfrag{spinUp}{\Huge $s=+1$}
\psfrag{spinDown}{\Huge $s=-1$}
\psfrag{nEq9}{\Huge $n=9$}
\psfrag{nEq0}{\Huge $n=0$}
\includegraphics[angle=-90,width=8.7cm]{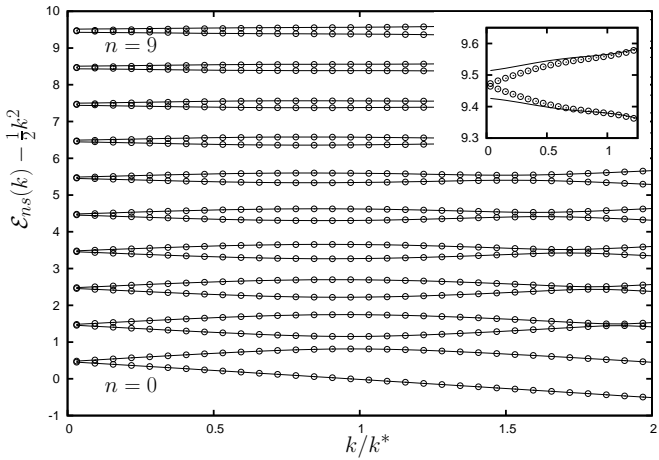}
\caption{Eigenenergies of the first ten transverse modes of the Rashba-Dresselhaus wire. Here the spin-orbit
coupling strength $k_\mathrm{RD}=0.25$ and
$\theta=0.126$ ($\alpha \gg \beta$) . The solid lines denote the approximate analytical solution in Eq.\ (\ref{eq:EnRD}), while the 
empty circles correspond to the numerical diagonalization of the corresponding Hamiltonian, Eq. (\ref{eq:HRDscaled}). 
The inset zooms in on the $n=9$ dispersion, showing more clearly the deviation between the analytical and numerical solutions.  The
maximum deviation is about $4\%$ near $k=0$.}
\label{fig:dispRD025}
\end{center}
\end{figure}
\begin{figure}[b]
\begin{center}
\psfrag{EAxis}{\Huge $\mathcal{E}_{ns}(k)-\frac{1}{2}k^2$}
\psfrag{kAxis}{\Huge $k/k^*$}
\psfrag{spinUp}{\Huge $s=+1$}
\psfrag{spinDown}{\Huge $s=-1$}
\psfrag{nEq9}{\Huge $n=9$}
\psfrag{nEq0}{\Huge $n=0$}
\includegraphics[angle=-90,width=8.7cm]{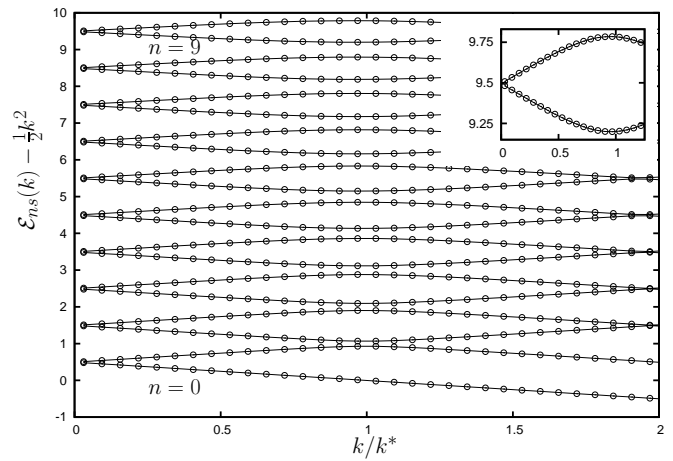}
\caption{Similar to Fig.~\ref{fig:dispRD025} with $k_\mathrm{RD}=0.10$ and
$\theta=0.126$ ($\alpha \gg \beta$). The deviation of the
analytical solution (lines) from the numerical (circles) is of the order of a few
percents, and on the scale of the plot can hardly be distinguished.}
\label{fig:dispRD010_LOW}
\end{center}
\end{figure}
\begin{figure}[t]
\begin{center}
\psfrag{EAxis}{\Huge $\mathcal{E}^0_{ns}(k)-\frac{1}{2}k^2$}
\psfrag{kAxis}{\Huge $k/k^*$}
\psfrag{spinUp}{\Huge $s=+1$}
\psfrag{spinDown}{\Huge $s=-1$}
\psfrag{nEq50}{\Huge $n=59$}
\psfrag{nEq59}{\Huge $n=50$}
\includegraphics[angle=-90,width=8.7cm]{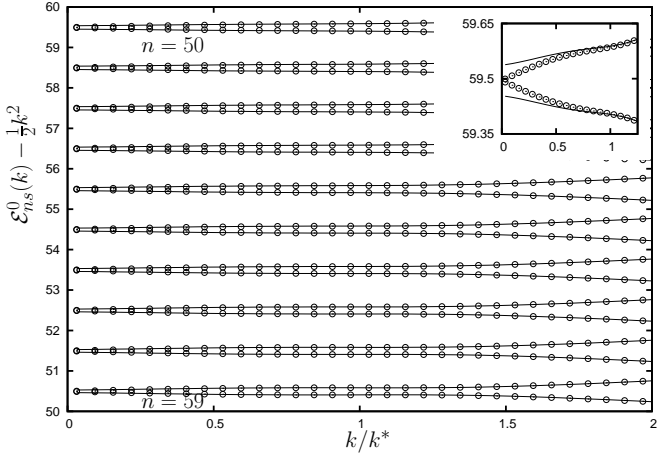}
\caption{Similar to Fig.~\ref{fig:dispRD010_LOW} but for the wire modes corresponding to $n=50-59$. 
Here  $k_\mathrm{RD}=0.10$ and $\theta=0.126$ ($\alpha \gg \beta$).
The inset shows more clearly the deviation
of the analytical solution (solid line) from the numerical one (circles) for the $n=59$ mode. The 
maximum deviation is less than $4\%$ near  $k=0$.  }
\label{fig:dispRD010}
\end{center}
\end{figure}
The coupling in Eq.\ (\ref{eq:HRDtilde}) is written in terms of the
exponential operators in order to remove all $\sigma_z$ contribution.  Thus, the
structure of the transverse coupling in the pure Rashba case is maintained,
i.e.\ only adjacent states with {\em opposite} spins are coupled.  By rewriting
the coupling
\begin{eqnarray}
e^{-ik_\mathrm{s}\hat{a}}(a^\dagger-a)e^{-ik_\mathrm{s}\hat{a}}&=&e^{-i\sqrt{2}k_\mathrm{s} a^\dagger} a^\dagger e^{-i\sqrt{2}k_\mathrm{s} a}
\nonumber \\
& &- e^{-i\sqrt{2}k_\mathrm{s} a^\dagger} a e^{-i\sqrt{2}k_\mathrm{s} a},
\end{eqnarray}
and writing the exponential operators as a power series one can show that the
diagonal matrix elements of the operator in the above equation is zero.
The exponential operators add the complication that states with opposite spins
separated by $N\hbar \omega$, $N \geq 2$, are coupled with coupling
strength $\propto k_\mathrm{s}^{N-1} =k_\mathrm{RD}^{2(N-1)}$, since we have assumed
that $\sin 2\theta \lesssim k_\mathrm{RD}$.  This assumption limits the possible
values of $\theta$  but the range of allowed values of $\theta$ can be extended
by repeating above the arguments around $|\alpha| \approx |\beta|$ and
$|\alpha |\ll |\beta|$, which will be shown in the next section.

\subsubsection{Effective Rashba-Dresselhaus wire Hamiltonian}

Now we can proceed as was done with Eq.\ (\ref{eq:HRprimeDefinition}) using the
same form of $S$ but using the substitution $k_\mathrm{R} \rightarrow
k_\mathrm{RD}$ in the $\sigma_z$ term and $k_\mathrm{R} \rightarrow k_\mathrm{c}$ in
the transverse coupling,
\begin{eqnarray}
H_\mathrm{RD}'&=&e^{S}\tilde{H}_\mathrm{RD}e^{-S}.
\label{eq:HRDprimeDefinition}
\end{eqnarray}
Using the same arguments as following Eq.\ (\ref{eq:HRprime}) we
obtain the effective Rashba and Dresselhaus Hamiltonian for the wire by discarding all
terms that lead to correction of order $(k_\mathrm{RD})^4$ :
\begin{eqnarray}
H_\mathrm{RD,eff}\!&\!=\!&\,\frac{k^2}{2}+ \hat{n}-k_\mathrm{RD} k
\,\sigma_z -\frac{(k_\mathrm{c})^2}{4}
\frac{1+(2\hat{n}+1)\sigma_z}{(1+2 k_\mathrm{RD} k)} \nonumber \\
& &\!\!\!+ \frac{1}{2}\bigl (\gamma_\mathrm{RD} a^\dagger \sigma_+ +a \sigma_-\gamma_\mathrm{RD}^\dagger
\bigr ) \nonumber  \\
& &\!\!\!+\frac{k_\mathrm{c}k_\mathrm{s}^2 }{2}(\sigma_+{a^\dagger}^2+\sigma_-a^2), 
\label{eq:HRDeff}
\end{eqnarray}
where $\gamma_{RD}$ is defined as
\begin{eqnarray}
\gamma_\mathrm{RD}&=&\frac{-i(k_\mathrm{c})^2}{\sqrt{2}}
 \left (
1- \frac{(k_\mathrm{c})^2 a^\dagger a}{4 (1 +2k_\mathrm{RD} k)} \right  ).
\label{eq:gammaRD}
\end{eqnarray}
Equations (\ref{eq:HRDeff}) and (\ref{eq:gammaRD}) have the same form as the
corresponding equations for the pure Rashba case, except for the $(\sigma_+
{a^\dagger}^2+\sigma_-a^2)$ term.

\subsubsection{Diagonalizing $H_\mathrm{RD,eff}$}

The goal now is to find a transformation that diagonalizes Eq.\ (\ref{eq:HRDeff}).
Instead of finding the exact transformation we will take advantage of the fact
that the rotation operator used in the pure Rashba case transforms the squared
ladder operator terms into an effective magnetic field term
$\mathcal{B}(\hat{n}) \sigma_x$ plus terms that can be treated perturbatively.
This proceedure is explained in Appendix \ref{app:a2rotation}.
The diagonal form of the Rashba and Dresselhaus Hamiltonian is thus
\begin{eqnarray}
H_\mathrm{RD;diag}&=& \mathcal{R}^\dagger \left (R^\dagger H_\mathrm{RD;eff}R
\right ) \mathcal{R} \nonumber \\
&=&\mathcal{R}^\dagger \left (
\begin{array}{cc}
H_+(\hat{n}) & \mathcal{B}(\hat{n})  \\
\mathcal{B}(\hat{n}) & H_-(\hat{n})
\end{array}
\right )\mathcal{R},
\label{eq:HRDdiag}
\end{eqnarray}
where $R$ is the same as in Eq.\ (\ref{eq:HReffRot}) with renormalized parameters [Eq.\ (\ref{eq:HRDeff})] and $\mathcal{R}$ is the operator 
that diagonalizes the $2\times2$ matrix (which commutes with $\hat{n}$) in Eq.\ (\ref{eq:HRDdiag}).
The quantity $\mathcal{B}(\hat{n})$ is given by
\begin{eqnarray}
 \mathcal{B}(\hat{n})=k_\mathrm{c}k_\mathrm{s}^2\frac{\sqrt{\hat{n}(\hat{n}+1)}}{2}(1-\cos\Theta
(\hat{n}) ),
\end{eqnarray}
see Appendices \ref{app:rotation} and \ref{app:a2rotation}, and
the eigenvalues of Eq.\ (\ref{eq:HRDdiag}) are
\begin{eqnarray}
\mathcal{E}_{n,\pm 1}&=&\frac{\varepsilon_{n,\uparrow}+\varepsilon_{n,\downarrow}}{2}
\pm \sqrt{\frac{(\varepsilon_{n,\uparrow}-\varepsilon_{n,\downarrow})^2}{4}
+\mathcal{B}^2(n)}. \label{eq:EnRD}
\end{eqnarray}
Note the form is just what one would expect from adding a magnetic field along the $x$-axis
to a system with a magnetic field, and quantization axis along $z$.

\subsubsection{Numerical vs analytical results: Rashba-Dresselhaus case}

Figures \ref{fig:dispRD025} - \ref{fig:dispRD010} show the comparison of the
analytical [Eq.\ (\ref{eq:EnRD})] and numerical eigenenergies of the Rashba-Dresselhaus wire, for different values of
$k_\mathrm{RD}$ and the maximum value of $\theta=0.126$, reflecting that
$\sin 2\theta \lesssim k_\mathrm{RD}$. The numerical eigenvalues are obtained by diagonalizing the 
Rashba-Dresselhaus wire Hamiltonian in Eq.\ (\ref{eq:HRDscaled}). The insets in Figs.\  \ref{fig:dispRD025}
and \ref{fig:dispRD010} zoom in on transverse states $n=9$ and $n=59$,
respectively.  On this scale the difference between the numerics (circles) and
the analytical solution (solid line) is more evident.  The maximum deviation,
closest to $k=0$, is of the order of $4\%$ for the highest values of $n$ but it
should be noted that the relative error if further reduced by a factor $\approx
59.5$.
Depending on the required accuracy, the analytical results may be extended beyond the strict requirement
that $\sin 2\theta \lesssim k_\mathrm{RD}$.

\subsection{Couplings $|\alpha| \ll |\beta|$ and $|\alpha| \approx |\beta|$}
\label{sec:alphaEQbeta}
The rotation introduced in Eq.\ (\ref{eq:HRDbar}) consists of two parts.  The
first part has the purpose of rotating the spin-orbit coupling term linear in
$k$ along $\sigma_y$, resulting in the same coupling as in the pure Rasbha case.
The second rotation simply transforms $\sigma_y\rightarrow \sigma_z$ and
does not affect the perturbation scheme itself.  The former rotation works for
any value of $\theta$, although the perturbation scheme for the transverse
coupling only works for $\sin 2\theta \leq k_\mathrm{RD}$.  With this in mind the
cases $|\alpha| \approx |\beta|$ and $|\alpha| \ll |\beta|$ can be treated in a
similar way.

When $|\alpha| \ll |\beta|$, it is convenient to parametrize the spin-orbit
angle as $\theta=\frac{\pi}{2}+\delta$, where $\delta$ is a small parameter.
The same steps are taken to arrive at the same equation as in (\ref{eq:HRDbar}) apart from the
new parameters $k_\mathrm{s}=k_\mathrm{RD} \sin(\pi+2\delta)=-k_\mathrm{RD}\sin(2\delta)$ and
$k_\mathrm{c}=k_\mathrm{RD}\cos(\pi+2\delta)=-k_\mathrm{RD} \cos(2\delta)$. Since the spectrum is even in the transverse coupling
parameters $k_\mathrm{c,s}$ the eigenenergies for $|\alpha| \ll |\beta|$ are the same as
those in Eqs.\ (\ref{eq:EnRD}).

If the couplings are of similar strength, $|\alpha| \approx |\beta|$, the angle
is parametrized as $\theta=\frac{\pi}{4}-\delta$, where $\delta$ is again a
small parameter.  The Hamiltonian after the rotation is
\begin{eqnarray}
\bar{H}_{\mathrm{RD}}&\equiv&   e^{-i\frac{\pi}{4}\sigma_x}e^{-\frac{i}{2} \left
( \frac{\pi}{4} -\delta\right )\sigma_z}
H_\mathrm{RD}e^{\frac{i}{2} \left ( \frac{\pi}{4} -\delta\right )
\sigma_z}e^{i\frac{\pi}{4}\sigma_x}  \nonumber \\
&=& \frac{1}{2}k^2+a^\dagger a
-k_\mathrm{RD} k \, \sigma_z  \nonumber \\
& &+\frac{i k_\mathrm{RD}}{\sqrt{2}}( \sin 2 \delta \sigma_x  -\cos 2\delta
\sigma_z)(a^\dagger -a),
\label{eq:HRDbarEQ}
\end{eqnarray}
which is identical to Eq.\ (\ref{eq:HRDbar}) apart from the switched roles of
\rm{sine} and \rm{cosine} due to the phase difference of $\pi/2$.
\begin{figure}[t]
\begin{center}
\psfrag{EAxis}{\Huge $\mathcal{E}^0_{ns}(k)-\frac{1}{2}k^2$}
\psfrag{kAxis}{\Huge $k/k^*$}
\psfrag{n9pBETA}{\Huge $\alpha \ll\beta \, :\, (9,\uparrow)$}
\psfrag{n9pEQ}{\Huge $\alpha \approx \beta\, :\, (9,\uparrow)$}
\psfrag{n5pBETA}{\Huge $\alpha \ll\beta\, :\, (5,\uparrow)$}
\psfrag{n5pEQ}{\Huge $\alpha \approx \beta\, :\, (5,\uparrow)$}
\includegraphics[angle=-90,width=8.7cm]{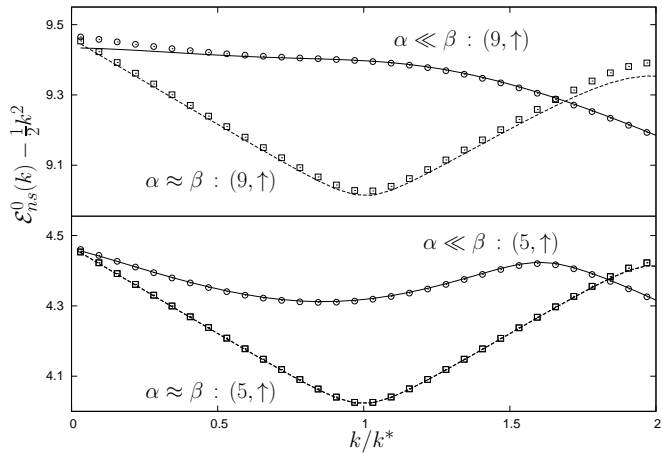}
\caption{Here we plot the numerical and analytical solution for $k_\mathrm{RD}=0.25$ with $\alpha \ll \beta$ (solid line and circles) and
$\alpha \approx \beta$ (dashed line and squares) for levels $(5,\uparrow)$ and $(9,\uparrow)$.
The analytical solution is the same as that derived for the $\alpha \ll \beta$ case.
Note that the $\alpha \approx \beta$ solution is accurate near $k=0$  but deviates more for $k\approx2k^*$, see text for details. }
\label{fig:dispRDCompare}
\end{center}
\end{figure}
As before, we can now proceed with proper unitary transformations in order to arrive at an effective
Hamiltonian, which can be similarly diagonalized to obtain an approximate analytical expression for the eigenvalues.

Figure \ref{fig:dispRDCompare} shows numerical and analytical results
for the eigenenergies of the Rashba-Dresselhaus wire with 
$k_\mathrm{RD}=0.25$ and different values of $\theta$:  $\theta=\pi/2-\delta$ and $\theta=\pi/4-\delta$, with $\delta=0.1$.
The solid line and the empty circles denote the $\theta=\pi/2-\delta$ results of the
analytical [Eq.\ (\ref{eq:EnRD})] and numerical solutions, respectively.  The
difference between the analytical and the numerical results are similar to the case with
$\theta=\delta$, i.e.\ they are close to 5\% corresponding the estimate given at
the end of Section \ref{sec:Rashba}.
The $\theta=\pi/4-\delta$ results look slightly different. The maximum
deviation between the analytical and the numerical results occurs near 
$k/k^*=2$, where the contribution of the $a^2$, ${a^\dagger}^2$ terms is the greatest.

This is best understood by looking at Eq.\ (\ref{eq:HRtilde}).  The phase shift
reverses definition of $k_\mathrm{s}$ and $k_\mathrm{c}$ so
$k_\mathrm{c}\propto k_\mathrm{RD}^2$ and $k_\mathrm{s}\propto k_\mathrm{RD}$.  Contributions from
higher order terms in the exponential are thus larger and lead to
the deviation around $k/k^*=2$.
This demonstrates that the perturbation scheme for $|\alpha|\ll |\beta|$ also
works, within the accuracy determined by $\delta$, by adjusting the value of the
spin-orbit angle according to $|\alpha | \approx |\beta|$ and $|\alpha| \gg |\beta|$.

\section{Magnetotransport in Rashba wires and effective $g$ factors in so coupled 2DEGs}
\label{sec:magnetotransport}
In this section we treat two relevant physical problems in which the approximation scheme
developed in the preceding sections can be applied: (i) magnetotransport in a Rashba wire and
(ii) the calculation of g-factors in spin-orbit coupled (Rashba-Dresselhaus) 2DEGs. In both (i) and (ii) we consider perpendicular magnetic fields.

\subsection{Shubnikov-de Haas oscillations}

In quantum wires the Shubnikov-de Haas oscillations in the magnetoresistance (along the wire) arise from the subsequent
depopulation of the transverse wire modes as the bottom of their bands cross the Fermi energy  $E_F$ for
increasing magnetic fields \cite{berggren88:10118}. This is similar to what happens in 2DEGs and also
in bulk metal systems [see Chapter 11 in Ref. \onlinecite{kittel87:book}]. Similarly to 2DEGs, in quantum wires
Shubnikov-de Haas oscillations allow the extraction of spin-orbit coupling strengths from the oscillation beating patterns. Following the
prescription in Refs.\ \onlinecite{knobbe05:035311} and \onlinecite{guzenko07:577}, 
below we outline in detail how magnetoresistance beating
patterns can emerge from the uneven distribution of the positions (in magnetic field) at which
the subsequent bottoms of the transverse bands cross $E_F$.

\begin{figure}[t]
\begin{center}
\psfrag{EAxis}{\Huge $\varepsilon_{ns}(k,B=0)$}
\psfrag{kAxis}{\Huge $k$}
\includegraphics[angle=-90,width=8.7cm]{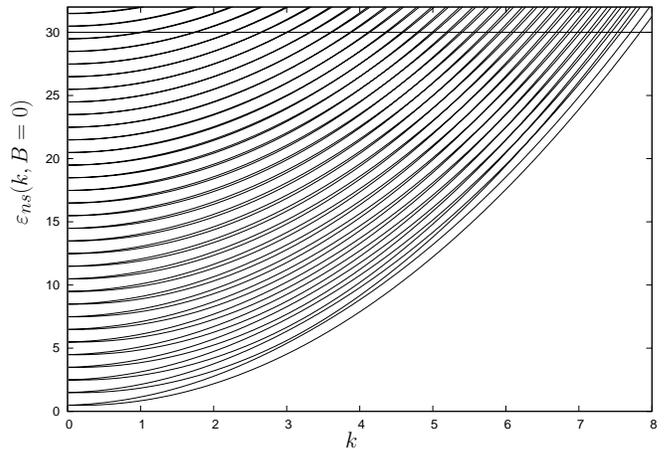}
\caption{$B=0$ eigenenergies of a Rashba wire [Eq.\ (\ref{eq:EnRD})] as a function of $k$ for a Rashba
coupling strength $k_\mathrm{R}=0.135$ and $E_F=30 \hbar
\omega_0$.}
\label{fig:Ensk}
\end{center}
\end{figure}

Figure \ref{fig:Ensk} shows a plot of the dispersion relation [from Eqs.\ (\ref{eq:EnU}) and (\ref{eq:EnD})] for a parabolic quantum
wire with Rashba spin-orbit coupling $k_\mathrm{R}=0.135$ in the {\em absence}
of a magnetic field. Note that here the plot includes the kinetic energy contribution.
Applying a magnetic field $B$ perpendicular to the 2DEG increases the effective
confinement of the wire, resulting in an effective confinement strength of
\begin{eqnarray}
\Omega=\sqrt{\omega_0^2+\omega_c^2},
\end{eqnarray}
where $\omega_c=eB/m^*$ is the cyclotron frequency.  This raises the bottom of the bands,
thus shifting the spectrum to higher energies.  As the magnetic field increases the
highest states below $E_F$ are pushed above $E_F$.  This gives rise to a peak in the magnetoresistance.
As the magnetic field is further increased more states cross $E_F$ thus leading to
oscillations in the Shubnikov-de Haas signal \cite{bastard88:book}.
As was discussed in Sec.\ \ref{sec:Rashba} the minima of $\varepsilon_{n,s}(k)$ occur at $k \approx 0$, at least for high enough $n$.
For lower values of $n$, where the minina is closer to $k\approx k_\mathrm{R}$, and the approximation $k\approx 0$ leads to an error in the
roots of order  $k_\mathrm{R}^2$.  Hence, further below when we need to determine the values of $B$
for which the highest energy level crosses $E_F$ we will use $\varepsilon_{ns}(k=0,B)=E_F$.\cite{knobbe05:035311,guzenko07:577}

Using the ladder representation the Hamiltonian of a quantum wire in a
perpendicular magnetic field at $k=0$ (wave number along the wire) is given by
\begin{eqnarray}
H&=&\sqrt{1+r^2}a^\dagger a-\frac{|g^*|}{4}\frac{m}{m_0}r \sigma_z +
\frac{ik_\mathrm{R+} }{2\sqrt{2}}(a^\dagger \sigma_+ -a \sigma_-)
\nonumber  \\
& &+ \frac{i k_\mathrm{R-}}{2\sqrt{2}}(a^\dagger \sigma_- -a \sigma_+),
\label{eq:HwqrB}
\end{eqnarray}
where $r=\omega_c/\omega_0$ and the bare electron mass $m_0$ appears through
$\frac{m}{m_0}=\frac{m\mu_B}{2e}$.  The coupling coefficients are defined as
\begin{eqnarray}
k_\mathrm{R+} &=&k_\mathrm{R} \sqrt{1+r^2}(1+r^2/(1+r^2)) \\
k_\mathrm{R-} &=&k_\mathrm{R} \sqrt{1+r^2}(1-r^2/(1+r^2)).
\end{eqnarray}
In writing Eq.\ (\ref{eq:HwqrB}) we assumed a negative $g$-factor.
By comparing Eqs.\ (\ref{eq:HwqrB}) and (\ref{eq:HRtilde}) we see that $g^*$ plays the
same role as $k$ in the $\sigma_z$ term.  The relation $\varepsilon_{n,s}(k<0)=\varepsilon_{n,-s}(k>0)$
gives the spectrum for the opposite signs of the $g$-factor by simply reversing the spin label, as
discussed following Eq.\ (\ref{eq:H0tilde}) on how to access the $k<0$ part of the spectrum.
Note that the coupling strengths above are not identical, since the magnetic field breaks time reversal
symmetry. This is in contrast to Eqs. (\ref{eq:W}) and (\ref{eq:V}). However, as we pointed out following Eq.\ (\ref{eq:V})
this does not affect the approximation scheme we have developed.
The eigenvalues are obtained by substituting the parameters in Eq.\  (\ref{eq:HwqrB}) into Eqs.
(\ref{eq:EnUx}) and (\ref{eq:EnDx}) in Appendix \ref{app:unequalVW}.

Figure \ref{fig:EnsB} shows the bottom of the energy bands $\varepsilon_{ns}(k=0,b)$ as a function of the magnetic field.  As the magnetic
field is increased the $k=0$ levels subsequently cross $E_F$, thus depopulating completely the corresponding transverse mode. 
These crossings occur at discrete values of $r$ that we denote by $r_i$, which correspond to magnetic field values $B_i=\frac{m^*\omega_0}{e} r_i$.
The black dots in the inset of Fig.\ \ref{fig:EnsB} shows the first 13 values of $r_i$ corresponding to solutions of 
$\varepsilon_{ns}(k=0,r_i)=E_F/\omega_0$ for different values of $(n,s)$.
\begin{figure}[t]
\begin{center}
\psfrag{EAxis}{\Huge $\varepsilon_{ns}(k=0,B)$}
\psfrag{BAxis}{\Huge $\omega_c /\omega_0$}
\includegraphics[angle=-90,width=8.7cm]{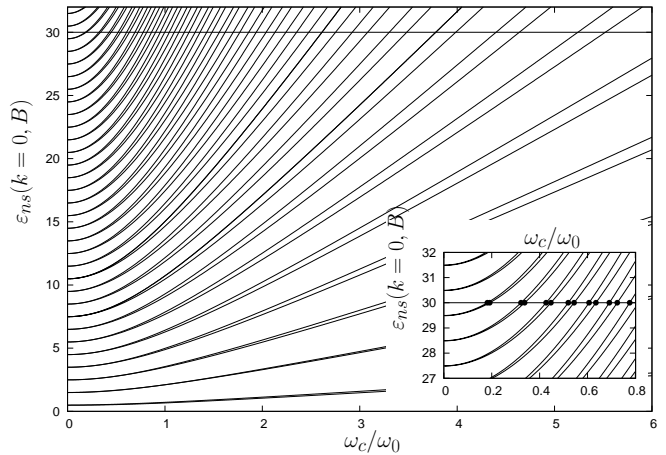}
\caption{Eigenenergies [Eqs.\
(\ref{eq:EnU}) and (\ref{eq:EnD})] as a function of the magnetic field with $k=0$, $k_\mathrm{RD}=0.135$, $g^*=-4$, $m/m_0=0.037$, and
$E_F=30 \hbar\omega_0$.  The inset shows a zoom of the intersections of the dispersion curves with $E_F$. }
\label{fig:EnsB}
\end{center}
\end{figure}
This results in a sequence of roots $\{r_i\}$, each corresponding to the specific magnetic field value $B_i=\frac{m^*\omega_0}{e} r_i$.

\begin{figure}[t]
\begin{center}
\psfrag{DrAxis}{\Huge $\Delta r_i$}
\psfrag{rAxis}{\Huge $r_i$}
\psfrag{0125}{\huge $k_\mathrm{R}=0.125$}
\psfrag{0135}{\huge $k_\mathrm{R}=0.135$}
\psfrag{0145}{\huge $k_\mathrm{R}=0.145$}
\includegraphics[angle=-90,width=8.7cm]{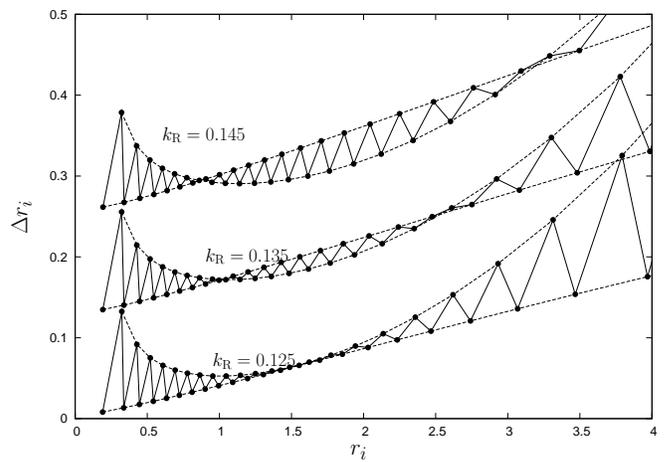}
\caption{Separation between subsequent roots $\Delta r_i=r_{i+1}-r_i$, plotted
as a function of $r_i$.  The separation of the node points, e.g., $r\approx 1.0$
and $r \approx 2.5$ for the uppermost curve, is determined by the value $\alpha/(\ell_0 \hbar \omega)$.}
\label{fig:SdeH}
\end{center}
\end{figure}

Plotting the difference between subsequent roots $\delta r_i=r_{i+1}-r_i$ as a function of $r_i$
gives rise to the data (black dots) in Fig.\ \ref{fig:SdeH}.
The dashed envelope in Fig.\ \ref{fig:SdeH} highlights the beating pattern of the crossings.  As discussed in Ref.\ \onlinecite{guzenko07:577} 
the data in Fig.\ \ref{fig:SdeH} is directly related to the  Shubnikov-de Haas oscillations of the wire: intuitively, the subsequent 
magnetic-field-induced depopulation of the transverse subbands affects the resistance of the wire because the number of conducting 
channels is being reduced. Hence the beating pattern describing the magnetic-field positions of the subsequent crossings of the band bottoms 
should manifest itself also in the magnetoresistance of the wire.

We emphasize that the observed beating pattern arises from the superposition of the two spin-dependent contributions that have different 
frequencies: the Zeeman splitting and the Rashba coupling.
The parameters used in the figure are $k_\mathrm{R}=0.125$, 0.135,
and 0.145,  $g^*=-4$, and $m/m_0=0.037$, corresponding to Ga$_{0.23}$In$_{0.77}$As.
Assuming $\hbar \omega_0=1$\,meV, the value $k_\mathrm{R}=0.135$ corresponds to a Rashba coupling
$\alpha=6.38$\,meV\,nm; both are realistic experimental values.\cite{knobbe05:035311,guzenko07:577}
By examining the Shubnikov-de Haas oscillations obtained in experiments the difference in magnetic field between the node points can be extracted. 
The value of the experimental node separation (in terms of magnetic field) can the used to extract the spin-orbit coupling from the
numerics shown in Fig. {\ref{fig:SdeH}}, i.e., finding $\alpha$ that will give the node point separation equal to the experimental value.

Since in our approach the determination of the relevant intersecting fields $B_i=\frac{m^*\omega_0}{e} r_i$ is straightforward 
via the transcendental equation $\varepsilon_{ns}(k=0,r_i)=E_F/\omega_0$), we believe the procedure just outlined should 
make it simpler to extract of the Rashba spin-orbit coupling from Shubnikov-de Haas oscillations in quantum wires.

\subsection{Landau-level $g$-factor}

Our approximation scheme can also be used to obtain the effective $g$-factor of a 2DEG in a
perpendicular magnetic field and in the presence of both Rashba and Dresselhaus interaction.
Again assuming a negative $g$-factor, the Hamiltonian of the system is
\begin{eqnarray}
H&=&\frac{1}{2m}(\Pi_x^2+\Pi_y^2)-\frac{1}{2}|g^*|\mu_B B
\sigma_z+\frac{\alpha}{\hbar}(\Pi_y \sigma_x-
\Pi_x \sigma_y) \nonumber \\
& &+\frac{\beta}{\hbar}(\Pi_x \sigma_x-\Pi_y  \sigma_y),
\end{eqnarray}
where $\bm{\Pi}=\bm{p}+\frac{eB}{2}\bm{r}\times  \hat{\bm{e}}_z$.
By defining the usual ladder operators
\begin{eqnarray}
a&=&\frac{\ell_c}{\sqrt{2}\hbar}(\Pi_x+i\Pi_y),
\end{eqnarray}
where $\ell_c=\sqrt{\hbar/eB}$, the Hamiltonian can be written as
\begin{eqnarray}
H&=&\hbar \omega_ca^\dagger a- \frac{|g^*|}{2}\mu_B B \sigma_z+\frac{
2 \beta}{2\sqrt{2}\ell_c}(a^\dagger \sigma_+ +a \sigma_-)
\nonumber \\
& &-\frac{i2\alpha}{2\sqrt{2}\ell_c}(a^\dagger \sigma_- - a \sigma_+).
\label{eq:LL_RD}
\end{eqnarray}
Through the transformation $a\rightarrow ae^{-i3\pi/4}$ and $\sigma_+ \rightarrow
\sigma_+e^{-i\pi/4}$ and measuring the energy in terms of of the cyclotron energy $\hbar \omega_c=\hbar eB/m^* $ we recover the Hamiltonian in
Eq.\ (\ref{eq:HRtilde}) with renormalized parameters.
Due to the magnetic field, the Hamiltonian in Eq.\ (\ref{eq:LL_RD}) is no longer
symmetric under time reversal and the Jaynes-Cummings term $W$ and the
perturbation term $V$ are no longer transformed into one another under time
reversal, but as mentioned before this does not affect the approximation scheme.

The energy spectrum of the Hamiltonian in Eq.\ (\ref{eq:LL_RD}) is obtained by
substituting the renormalized coupling parameters into Eqs.\ (\ref{eq:EnUx})-(\ref{eq:Deltax}), in Appendix \ref{app:unequalVW}.
This shows that the Hamiltonian for a 2DEG in a perpendicular magnetic field
in the presence of Rashba and Dresselhaus couplings is formally equivalent to a
parabolically confined 2DEG with Rashba interaction.
Note that the spectrum for pure Dresselhaus coupling, $\alpha=0$, reduces to the known result
for 2DEG in a perpendicular magnetic field and with Dresselhaus coupling \cite{winkler03:1}.

From the energy spectrum$\varepsilon_{ns}$ the $g$-factor
can be defined for different Landau levels.  We do not include the effects of electron-electron interaction, e.g.\ exchange enhancement of
the $g$-factor \cite{nicholas88:1294}, but rather focus on the effect of the spin-orbit coupling.
The $g$-factor of Landau level $n$ is defined as the difference of Kramers doublets divided by $\mu_B B$
\begin{eqnarray}
g_{n}=\frac{\varepsilon_{n\uparrow}-\varepsilon_{n\downarrow}}{\mu_B B}.
\label{eq:gLL}
\end{eqnarray}
In the absence of spin-orbit coupling the $g$-factor is equal to $g^*$ for all Landau levels.

\begin{figure}[t]
\begin{center}
\psfrag{gFactor}{\Huge $g_n$}
\psfrag{alpha}{\Huge $\alpha$\,[meV\,nm]}
\psfrag{gEff1-0}{\huge $|g^*|=1,\,n=0$}
\psfrag{gEff1-3}{\huge $|g^*|=1,\,n=3$}
\psfrag{gEff4-0}{\huge $|g^*|=4,\,n=0$}
\psfrag{gEff4-3}{\huge $|g^*|=4,\,n=3$}
\includegraphics[angle=-90,width=8.7cm]{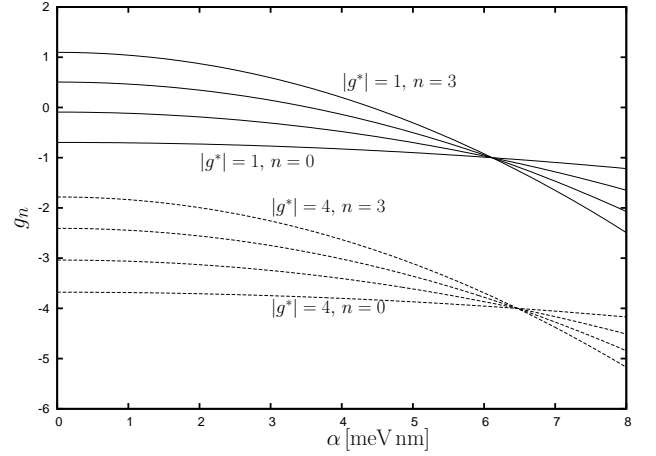}
\caption{$n^{th}$-Landau-level $g_n$-factors for two different values of $|g^*|=4$ and $|g^*|=1$, as a function of Rashba coupling.  Other parameter values are $B=2$\ T, $\frac{m}{m_0}=0.037$, and $\beta=6$\,meV \,nm.}
\label{fig:gFactor}
\end{center}
\end{figure}

In Fig.\ \ref{fig:gFactor} we plot the effective $g$-factor for the first four
Landau levels for two values of $|g^*|$.  The parameter values are
$B=2$\ T, $\frac{m}{m_0}=0.037$, and $\beta=6$\,meV \,nm and
different $g$-factor values $|g^*|=4$ and $|g^*|=1$, while $\alpha$ ranges from zero to 8\,meV \,nm.

For the lower value $|g^*|=1$ the effective $g$-factor changes sign for $n\geq 2$
for low values of $\alpha$.  As $\alpha$ is raised all the $g_n$'s cross at a single
value of $\alpha$.  Similar crossing behavior is observed for $|g^*|=4$ but the sing of $g_n$
remains the same since for $n=0-3$.
The results shown in Fig. \ref{fig:gFactor} are best understood by expanding Eq.\ (\ref{eq:gLL}) assuming
that $\alpha, \beta \ll \hbar \omega_c \ell_c /2$. The $g$-factor then becomes
\begin{eqnarray}
\frac{g_{n}}{|g^*|}\approx-1+\frac{8(n+\frac{1}{2})}{\frac{|g^*|}{2}\frac{m}{m_0}}\left (
\frac{(k_\mathrm{D}\ell_c)^2}{1-\frac{|g^*|}{2}\frac{m}{m_0}}-\frac{(k_\mathrm{R}\ell_c)^2}{1+\frac{|g^*|}{2}\frac{m}{m_0}}
\right ).
\label{eq:gLLexpansion}
\end{eqnarray}
Note that $\frac{|g^*|}{2}\frac{m}{m_0}\ll 1$ for typical parameters. The above equation shows two important things.  First, the $g$-factor can change sign when the spin-orbit part exceeds $-|g^*|$
which can occur for high enough values of $n$ or $\beta$.  Secondly, the effect of the spin-orbit interaction is completely canceled when the Rashba and the Dresselhaus couplings satisfy the condition
\begin{eqnarray}
(k_\mathrm{D}\ell_c)^2&=&(k_\mathrm{R}\ell_c)^2\frac{1-\frac{|g^*|}{2}\frac{m}{m_0}}{1+\frac{|g^*|}{2}\frac{m}{m_0}}.
\label{eq:gNoSO}
\end{eqnarray}
Such crossings are a general feature of having both the Rashba and the Dresselhaus couplings but the sign change can only occur for small enough values of $|g^*|$.\cite{zarea05:085342}
Note that if $g^*=0$, this condition corresponds to $\alpha=\pm \beta$, where the energy spectrum is known and the spin splitting of Kramers doublets is zero. The above results may prove useful for experimentalist to extract spin-orbit coupling strengths via g-factor measurements.

\section{Conclusion}
We have developed an analytical approximation scheme suitable for obtaining the energy spectrum of systems with both the Rashba
and Dresselhaus spin-orbit couplings and parabolic confinement, e.g., due to a magnetic field or electrostatic gating.
We have applied our approach to investigate (i) magnetotransport in quantum wires
with the Rashba coupling and (ii) the effective $g$-factor of a 2DEG in a perpendicular magnetic field in the
presence of both the Rashba and the Dresselhaus couplings. In (i) our approximate eigenvalues
allow us to obtain beating patterns that relate to the Shubnikov-de Haas oscillations, by simply finding the roots of
transcendental equations.  
For (ii) we derive a relation for the Landau level dependent effective $g_n$-factor.
We find that for small enough bulk $g$-factors the spin-orbit effects drop out of $g_n$ for certain values of the Rashba
and Dresselhaus couplings. We believe our analytical approach should allow for a more straightforward extraction of spin-orbit 
coupling strengths from Shubnikov-de Haas oscillations in wires and g-factor measurements in 2DEGs.

\section{Acknowledgement}
This work was supported by the Icelandic Research Fund, the Icelandic Science
and Technology Research Programme for Postgenomic Biomedicine, Nanoscience and
Nanotechnology, the Swiss NSF, the NCCR Nanoscience, DARPA, CNPq and FAPESP.

\appendix
\section{Rotation of generalized Jaynes-Cummings}
\label{app:rotation}
The Hamiltonians in Eqs.\ (\ref{eq:HReff}) and (\ref{eq:HRDeff}) can be written
in the form
\begin{eqnarray}
H_\mathrm{eff}&=&
\left (
\begin{array}{cc}
h_1(\hat{n},k) & \gamma(\hat{n},k) a^\dagger \\
  a\gamma^*(\hat{n},k)  & h_2(\hat{n},k)
\end{array} \right ),
\label{eq:Heff}
\end{eqnarray}
where $\hat{n}=a^\dagger a$ and the diagonal elements are defined as
$h_1(a^\dagger a,k)=\langle
+1|H_\mathrm{eff}|+1\rangle$ and $h_2(a^\dagger a,k)=\langle
-1|H_\mathrm{eff}|-1\rangle$.  All the matrix elements depend on $k$ but since
it is only a parameter
it is dropped for convenience.
The goal is to find a unitary transformation, or rotation, $R$ that diagonlizes
the Hamiltonian in Eq.\ (\ref{eq:Heff}). If the ladder operator $a\,
(a^\dagger)$ is replaced by a complex number $\alpha\, (\alpha^*)$ the rotation
operator reduces to
\begin{eqnarray}
  R_0&=&
\left (
\begin{array}{cc}
\cos \frac{\theta}{2} & \frac{\gamma^* \alpha}{|\gamma|\sqrt{\alpha^*
\alpha}}\sin \frac{\theta}{2} \\
-\frac{\gamma \alpha^*}{|\gamma|\sqrt{\alpha^* \alpha}}\sin \frac{\theta}{2}  &
\cos \frac{\theta}{2}
\end{array} \right ).
\label{eq:R0cosine}
\end{eqnarray}
Note that the factor $ \frac{\gamma \alpha}{|\gamma | \sqrt{\alpha^* \alpha}}$
can equivalently be written in the ususal phase factor form $e^{i\phi}$.  The
choice of our notation will help to understand the form of the general rotation
matrix discussed below, see Eq.\ (\ref{eq:Rcosine}).
To take into account the commutation properties of the ladder operators in the
diagonalization of Eq.\ (\ref{eq:Heff}), we start with the form
\begin{eqnarray}
 R&=&\left (
\begin{array}{cc}
\frac{1}{\sqrt{1+X^\dagger X}} & X^\dagger \frac{1}{\sqrt{1+X X^\dagger}}  \\
\frac{-1}{\sqrt{1+X X^\dagger}}X & \frac{1}{\sqrt{1+X X^\dagger}}
\end{array} \right ),
\end{eqnarray}
where $X$ is an operator that depends only on the ladder operator, to be
determined. The diagonalization requires that the $2\times 2$ spin matrix
\begin{eqnarray}
H_d=R^\dagger H_\mathrm{eff} R
\label{eq:Hd}
\end{eqnarray}
has zero off-diagonal elements.  Using the {\em ansatz} $X^\dagger=-a^\dagger
\chi(\hat{n})$ and the relation $f(\hat{n})a =a f(\hat{n}-1)$
the diagonalization is achieved when
$\chi(\hat{n})=\frac{1}{\sqrt{\hat{n}+1}}\tan \frac{\Theta(\hat{n})}{2}$.
and the rotation operators takes the form
\begin{eqnarray}
 R&=&\left (
\begin{array}{cc}
(1-\mathcal{Q}_0)+\cos\frac{\Theta(\hat{n}-1)}{2}\mathcal{Q}_0 &
-\frac{\gamma(\hat{n})}{|\gamma(\hat{n})|} a^\dagger \frac{\sin
\frac{\Theta(\hat{n})}{2}}{\sqrt{\hat{n}+1}}  \\
\frac{\sin \frac{\Theta(\hat{n})}{2}}{\sqrt{\hat{n}+1}} a
\frac{\gamma^*(\hat{n})}{|\gamma(\hat{n})|} &
\cos\frac{\Theta(\hat{n})}{2}
\end{array} \right ). \nonumber  \\
\label{eq:Rcosine}
\end{eqnarray}
The diagonalization is achieved when
$\chi(\hat{n})=\frac{1}{\sqrt{\hat{n+1}}}\tan \frac{\Theta(\hat{n})}{2}$.
Here we have introduced the operator $\mathcal{Q}_0=a^\dagger \frac{1}{ a^\dagger
a+1}a$.
The operator $\mathcal{Q}_0$ acts as the complement of the projector into the
oscillator ground-state $\mathcal{P}_0=|0 \rangle \langle 0 |$, i.e.\
$\mathcal{Q}_0|n \rangle=|n\rangle$ if $n>0$ and  $\mathcal{Q}_0|0 \rangle=0$.
It does not appear in any of the other matrix elements since only the operator
$1/\sqrt{1+X^\dagger X}$ gives rise to $\mathcal{Q}_0$, not the other form
$1/\sqrt{1+X X^\dagger}$.  This asymmetry comes from the fact that
$|0,+1\rangle$ is an eigenstate of Eq.\ (\ref{eq:Heff}) but not of $|0,-1\rangle$
or of any other state $| n>0,s \rangle$.
The angle operator $\Theta(\hat{n})$ is defined via the ususal relation
\begin{eqnarray}
\cos \Theta(\hat{n})=\frac{h_1(\hat{n}+1)-h_2(\hat{n})}{\sqrt{(h_1(\hat{n}
+1)-h_2(\hat{n}))^2 + 4(\hat{n}+1)  |\gamma(\hat{n}+1 )|^2}}. \nonumber \\
\label{eq:CosineOperator}
\end{eqnarray}
The form of the matrix in Eq.\ (\ref{eq:Rcosine}) is reminiscent of the form of
Eq.\ (\ref{eq:R0cosine}), the difference coming from the non-commutivity of $a$
and $a^\dagger$ and the operator $\mathcal{Q}_0$.

The eigenenergies are determined by the diagonal elements of $H_\mathrm{d}$,
which are given by
\begin{eqnarray}
{[H_{\mathrm{d}}]}_{2,2}&=&\frac{1}{2} \Bigl  ( (h_1(\hat{n}+1)-h_2(\hat{n}))
\nonumber \\
& &  -\sqrt{(h_1(\hat{n}+1)-h_2(\hat{n}))^2+4(\hat{n}+1) |\gamma(\hat{n}+1)|^2}
\Bigr ) \nonumber \\
\label{eq:Hd11} \\
{[H_{\mathrm{d}}]}_{1,1}&=&h_1(\hat{n})(1-\mathcal{Q}_0)+ \frac{1}{2}\Bigl (
(h_1(\hat{n})-h_2(\hat{n}-1)) \nonumber  \\
& &+\sqrt{(h_1(\hat{n})-h_2(\hat{n}-1))^2+4\hat{n} |\gamma(\hat{n})|^2} \Bigr
)\mathcal{Q}_0. \nonumber \\
\label{eq:Hd22}
\end{eqnarray}
The two equations above are obtained by algebraic manipulations of Eq.\
(\ref{eq:Hd}) and collecting all powers of $a$ from $1/\sqrt{1+X^\dagger X}$ and
writing them in terms of $\mathcal{Q}_0\equiv a^\dagger \frac{1}{a^\dagger a +1}a$
and $1-\mathcal{Q}_0$.

\section{Rotation of $\sigma_+ {a^\dagger}^2+ \mbox{h.c.\ }$}
\label{app:a2rotation}
The effective Hamiltonian in Eq.\ (\ref{eq:HRDeff}) cannot be exactly
diagonalized using the rotation in App.\ \ref{app:rotation}.  This is due to the term
$\frac{1}{2}(\sigma_+{a^\dagger}^2+\mathrm{h.c.\ })$.  From the Hamiltonian that results from
the rotation it is easy to discard terms which give
corrections $\mathcal{O}((k_\mathrm{RD})^4)$.
Using the property $\frac{\gamma^*(\hat{n})}{|\gamma(\hat{n})|}=i$, the
transformed second order coupling is
\begin{eqnarray}
& &R^\dagger \frac{1}{2}(\sigma_+{a^\dagger}^2+\sigma_-a^2)R \nonumber \\
 &=&\sqrt{\hat{n}(\hat{n}+1)}\sin\frac{\Theta(\hat{n}-1)}{2}\sin\frac{
\Theta(\hat{n})}{2}\sigma_x \nonumber \\
& &+i\sigma_z \left ( a^\dagger
\sin\frac{\Theta(\hat{n}-1)}{2}\cos\frac{\Theta(\hat{n})}{2} -\mbox{h.c.\ }
\right ) \nonumber \\
& & +  \frac{1}{2} \left ( \sigma_+ \left (  a^\dagger \cos
\frac{\Theta(\hat{n})}{2} \right )^2 +\mbox{h.c.\ } \right ).
\label{eq:a2rotation}
\end{eqnarray}
As can be seen in Figs.\ \ref{fig:dispRashba025} and
\ref{fig:dispRashba010}, the only states that cross are the same transverse
states with {\em opposite} spin, e.g.\ when $n=1$ the $\uparrow$ and
$\downarrow$ states cross at $k \approx 3.75$ in Fig.\ \ref{fig:dispRashba025}.
Taking advantage of this behavior we can determine which terms in Eq.\
(\ref{eq:a2rotation}) need to be kept and which can be dropped.    The first
term in Eq.\ (\ref{eq:a2rotation}) couples states with with opposite spin but
same $n$ and the latter two terms couple {\em adjacent} transverse states that do not cross and can
be treated perturbatively.  They give rise to corrections in energy
$\mathcal{O}((k_\mathrm{RD})^4)$ and can be discarded.
From these considerations the form of the rotated second-order coupling is
\begin{eqnarray}
& &R^\dagger \frac{k_\mathrm{c} k_\mathrm{s}^2}{2}(\sigma_+{a^\dagger}^2+\sigma_-a^2)R \\
&=&k_\mathrm{c} k_\mathrm{s}^2\frac{\sqrt{\hat{n}(\hat{n}+1)}}{2}(1-\cos\Theta (\hat{n})
)\sigma_x+\mathcal{O}(k_\mathrm{c}^2k_\mathrm{s}^3). \nonumber
\end{eqnarray}

\section{The eigenenergies for general coupling values}
\label{app:unequalVW}
For completeness here we provide the eigenvalues of a generic Hamiltonian of the form
\begin{eqnarray}
H&=&\frac{k^2}{2}+a^\dagger a-X\sigma_z +i\frac{Y}{2\sqrt{2}}(a^\dagger \sigma_+- a \sigma_-) \\ \nonumber
& &+i\frac{Z}{2\sqrt{2}}(a^\dagger \sigma_-- a \sigma_+),
\end{eqnarray}
where $X, Y$ and $Z$ are real numbers and the last term, proportional to $Z$, is the perturbative coupling.
The only restriction on the parameter values are that $X>0$ and $Y$ and $Z$ should be of the same same
order of magnitude although their numerical value can be different.
This type of Hamiltonian covers all models discussed in this paper. Up to corrections of the order $Z^4$, the resulting eigenvalues are
\begin{eqnarray}
\varepsilon_{n,\uparrow}(k)&=&
\frac{k^2}{2}+n-\frac{Z^2/2}{1+2X}
+\Delta_{n} ,
\label{eq:EnUx}\\
\varepsilon_{n,\downarrow}(k)&=&
\frac{k^2}{2}+n+1-\frac{Z^2/2}{1+2X}
-\Delta_{n+1},
\label{eq:EnDx}
\end{eqnarray}
where
\begin{eqnarray}
\Delta_n&=&\frac{1}{2}\left ( \left ( 1 -2X-\frac{Z^2n}{1+2X} \right )^2 \right . \nonumber \\
 & & \left .+ 2 Y^2 n\left ( 1-\frac{Z^2 n}{4(1+2X)}\right )^2 \right )^{1/2}.
\label{eq:Deltax}
\end{eqnarray}
When $Z=0$ the spectrum reduces to the known Jaynes-Cummings solution.

%


\end{document}